\documentclass[aps, 
	prl, 
	noeprint, 
	superscriptaddress, 
	floatfix, 
	tightenlines, 
	twocolumn
	]{revtex4-1}
\usepackage[english]{babel}
\usepackage{graphicx}                      
\usepackage{listings} 
\usepackage{amsmath,amssymb,amsthm}        
\usepackage[utf8]{inputenc}
\usepackage[toc,page]{appendix}
\usepackage{color}
\usepackage{caption}

\usepackage{hyperref}
\hypersetup{
  colorlinks   = true, 
  urlcolor     = blue, 
  linkcolor    = blue, 
  citecolor    = blue 
}

\newcommand{\YSO}{Y$_2$SiO$_5$}
\newcommand{\PrYSO}{Pr$^{3+}$:Y$_2$SiO$_5$}

\newcommand{\gafc}{$g^{(2)}_\textit{AFC,i}$}

\newcommand{\ket}[1]{\left | #1 \right \rangle}	
\newcommand{\tx}[1]{\textup{#1}}		

\begin{document} 

\title{Storage and analysis of light-matter entanglement in a fibre-integrated system}
\author{Jelena V. Rakonjac}
\thanks{These authors contributed equally.}
\affiliation{ICFO - Institut de Ciencies Fotoniques, The Barcelona Institute of Science and Technology, 08860 Castelldefels (Barcelona), Spain.}
\author{Giacomo Corrielli}
\thanks{These authors contributed equally.}
\affiliation{Istituto di Fotonica e Nanotecnologie (IFN) - CNR, P.zza Leonardo da Vinci 32, 20133 Milano, Italy.}
\author{Dario Lago-Rivera}
\thanks{These authors contributed equally.}
\affiliation{ICFO - Institut de Ciencies Fotoniques, The Barcelona Institute of Science and Technology, 08860 Castelldefels (Barcelona), Spain.}
\author{Alessandro Seri}
\affiliation{ICFO - Institut de Ciencies Fotoniques, The Barcelona Institute of Science and Technology, 08860 Castelldefels (Barcelona), Spain.}
\author{Margherita Mazzera}
\affiliation{Institute of Photonics and Quantum Sciences, SUPA, Heriot-Watt University, Edinburgh EH14 4AS, UK.}
\author{Samuele Grandi}\email[]{samuele.grandi@icfo.eu}
\affiliation{ICFO - Institut de Ciencies Fotoniques, The Barcelona Institute of Science and Technology, 08860 Castelldefels (Barcelona), Spain.}
\author{Roberto Osellame}
\affiliation{Istituto di Fotonica e Nanotecnologie (IFN) - CNR, P.zza Leonardo da Vinci 32, 20133 Milano, Italy.}
\author{Hugues de Riedmatten}
\affiliation{ICFO - Institut de Ciencies Fotoniques, The Barcelona Institute of Science and Technology, 08860 Castelldefels (Barcelona), Spain.}
\affiliation{ICREA - Instituci\'o Catalana de Recerca i Estudis Avan\c cats, 08015 Barcelona, Spain.}

\begin{abstract}
The deployment of a fully-fledged quantum internet poses the challenge of finding adequate building-blocks for entanglement distribution between remote quantum nodes. An ideal system would combine propagation in optical fibres with quantum memories for light, leveraging on the existing communication network while featuring the scalability required to extend to network sizes. Here we demonstrate a fiber-integrated quantum memory entangled with a photon at telecommunication wavelength. The storage device is based on a fiber-pigtailed laser written waveguide in a rare-earth doped solid and allows an all-fiber stable adressing of the memory. The analysis of the entanglement is performed using fibre-based interferometers. Our results feature orders of magnitude advances in terms of storage time and efficiency for integrated storage of light-matter entanglement, and constitute a significant step forward towards quantum networks using integrated devices.
\end{abstract}

\maketitle


\section{Introduction}
An interconnected set of quantum hubs will open the way to the next generation of quantum technology applications, from secure communication to distributed quantum computing \cite{Kimble2008, Wehner2018}. The basic ingredient for the realisation of these quantum networks is entanglement between light and matter \cite{Duan2001}, distributed over long distances exploiting quantum repeater architectures \cite{Simon2007, Sangouard2011}. However, requirements on fidelity and communication rate and distance place stringent conditions on the physical implementations, which should allow for a high degree of multiplexing and a clear path to scalability. Integrated systems, with a low spatial footprint and enhanced light-matter interaction, provide a promising route forward, and already provided enhanced performances in atomic systems \cite{Corzo2019}, quantum dots \cite{Uppu2020,You2021} and colour centres \cite{Sipahigil2016,Bhaskar2020}.

Rare-earth based systems \cite{Macfarlane2002} are excellent candidates for engineering light-matter interactions. Entanglement has already been stored and generated in these systems \cite{Clausen2011, Usmani2012, Kutluer2019, Lago-Rivera2021, Liu2021}, where photons are mapped into delocalised excitations of billions of ions. These quantum memories support multimode storage \cite{Seri2017, Yang2018, Seri2019}, and pairing them with an external entanglement source provides a bridge to telecommunication networks \cite{Lago-Rivera2021, Rakonjac2021a}. Alternatively, direct storage at telecommunication wavelengths is also possible \cite{Saglamyurek2015, Craiciu2021}. The solid-state nature of rare-earth based memories facilitates integration in photonic devices \cite{Corrielli2016, Zhong2017, Seri2018, Liu2020a}, leading to single-ion detection \cite{Dibos2018,Kindem2020}, and storage of entanglement in an integrated fashion \cite{Saglamyurek2011, Saglamyurek2015}.

In this direction, a final key step is to find an efficient way of coupling light in and out of the integrated device. Possible approaches include directly doping an optical fibre with rare-earth ions \cite{Saglamyurek2015}, where reaching good storage performances is challenging, or to adopt free-space coupling \cite{Saglamyurek2011, Zhong2017}, limiting the advantages of an integrated device. Alternatively, an integrated structure could be fabricated with a mode that is directly compatible with an optical fibre, which is then attached to the device.

Here we present a fibre-integrated solid-state system for light-matter entanglement storage. Our solution is based on a laser-written waveguide in a \YSO\ crystal doped with praseodymium (Pr) ions \cite{Seri2018}, which is directly coupled and permanently attached (pigtailed) to a single mode optical fibre. The mode of the waveguide is compatible with that of the fibre, providing an efficient and practical coupling between the integrated device and the optical fibre, strongly reducing the need for alignment of our optical setup and opening the way to exploiting the resources available to fibre-based and chip-based devices.

\section{Main}
\begin{figure*}
	\centering
	\includegraphics[width=2\columnwidth]{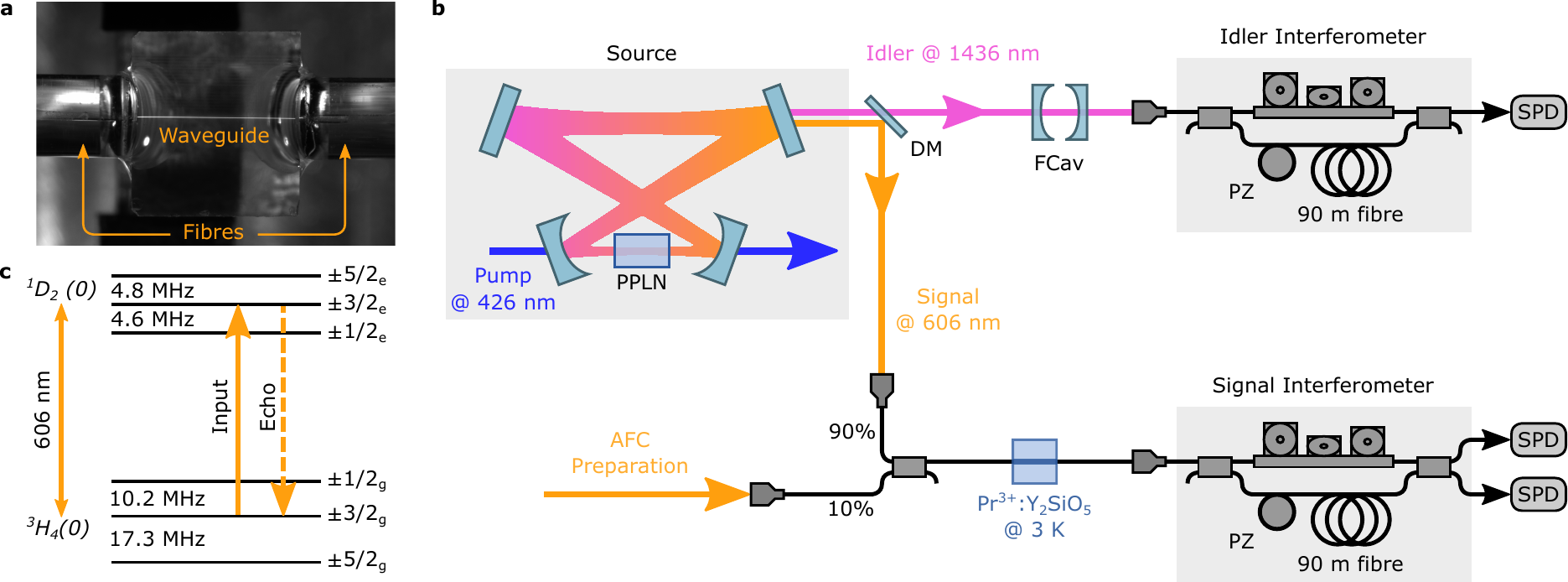}
	\caption{Experimental Setup. (a) Picture of the fibre-pigtailed waveguide. The bright line is the fluorescence from the Pr ions, excited by 607 nm light coupled into the waveguide. (b) Schematics of the setup. Entangled photon pairs are generated in a cavity-enhanced spontaneous parametric down-conversion (SPDC) source. The idler and the signal are separated by a dichroic mirror (DM), and the latter is coupled to the single mode fibre that is directly glued to the waveguide in the \PrYSO\ crystal, placed inside a cryostat. An AFC of varying storage time is prepared in the memory. The entangled state is analysed using two fibre-based unbalanced Mach-Zehnder interferometers, with the one for the signal photons directly linked to the waveguide. SPD: single photon detector. The signal photons are detected using silicon photodiodes, while superconducting detectors are used for idler photons. (c) Level scheme of \PrYSO.}
	\label{fig:1}
\end{figure*}
A picture of the fibre-pigtailed crystal is shown in Fig.~\ref{fig:1}(a). A type-I waveguide is written in the \PrYSO\ crystal using femtosecond laser micro-machining \cite{Seri2018}. Two single-mode optical fibres are then glued to the facets of the crystal to couple light in and out of the waveguide, with a total transmission of 25\%\ \cite{SuppMat}. Our experimental setup is sketched in Fig.~\ref{fig:1}(b). Energy-time entangled pairs of telecom ({\itshape idler}) and visible ({\itshape signal}) photons are generated in a cavity-enhanced spontaneous parametric down-conversion (SPDC) source \cite{Seri2018}, designed to emit narrow photons (1.8 MHz) compatible with light storage in Praseodymium.
Signal photons are coupled into a single-mode fibre, and directly routed to the fibre-coupled waveguide. The photons are then stored as a collective excitation of the Pr ions for a predetermined storage time using the Atomic Frequency Comb (AFC) protocol \cite{Afzelius2009}, obtained by shaping the inhomogeneously broadened profile of the $^3$H$_4$(0) $\leftrightarrow$ $^1$D$_2$(0) transition of the Pr ions into a periodic structure (Fig.~\ref{fig:1}(c)). The frequency period $\Delta$ of the comb determines the storage time of the AFC, as $\tau_\tx{AFC} = 1/\Delta$. The AFC preparation laser is also guided to the waveguide via the optical fibre , and mixed with the signal photons with a 90:10 fibre beamsplitter. The fibre-coupled crystal is placed inside a closed-cycle cryostat, and cooled to below 3 K. The fibre-to-fibre transmission is maintained at low temperatures and through several cooling cycles without significant deterioration.

\begin{figure}
	\centering
	\includegraphics[width=\columnwidth]{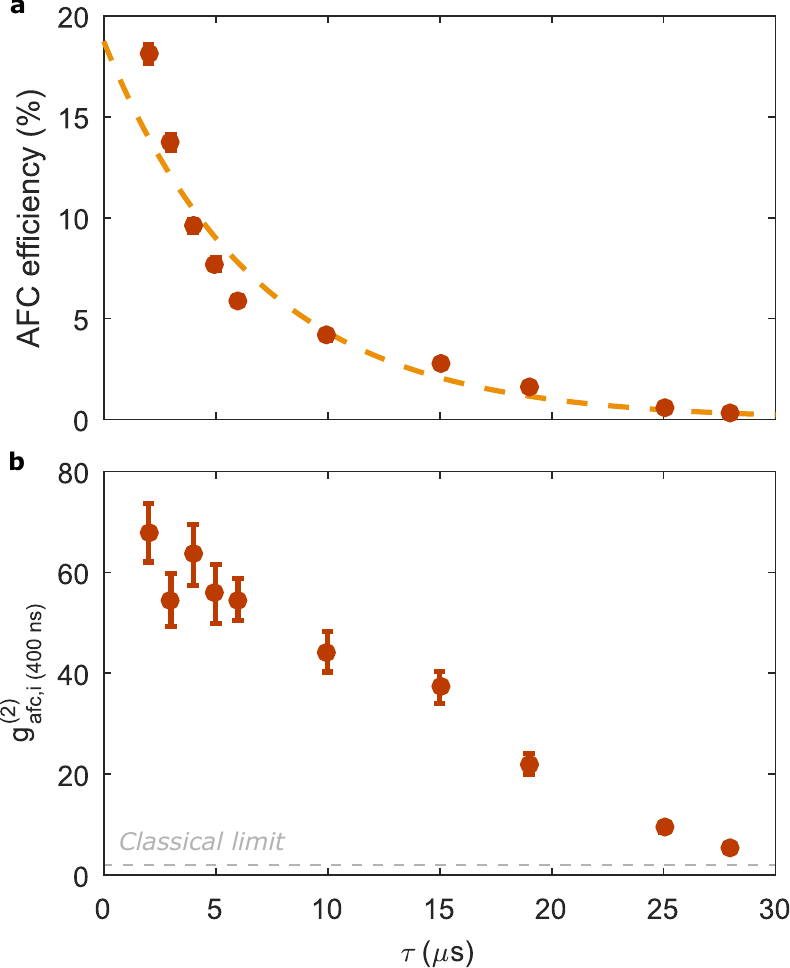}
	\caption{System characterisation at different storage times. (a) Efficiency of the AFC for increasing storage time, calculated with single photons in a 400 ns window. The dashed line is a fit to an exponential, from which we were able to extract an effective coherence time $T^\tx{eff}_2$ of 27(3) $\mu$s. (b) Second-order cross-correlation function \gafc\ between idler photons and retrieved AFC echo for increasing storage time. We measured non-classical correlations up to the longest storage time we analysed, 28 $\mu$s. For all the plots the error bars correspond to one standard deviation.}
	\label{fig:2}
\end{figure}
\begin{figure}
	\centering
	\includegraphics[width=\columnwidth]{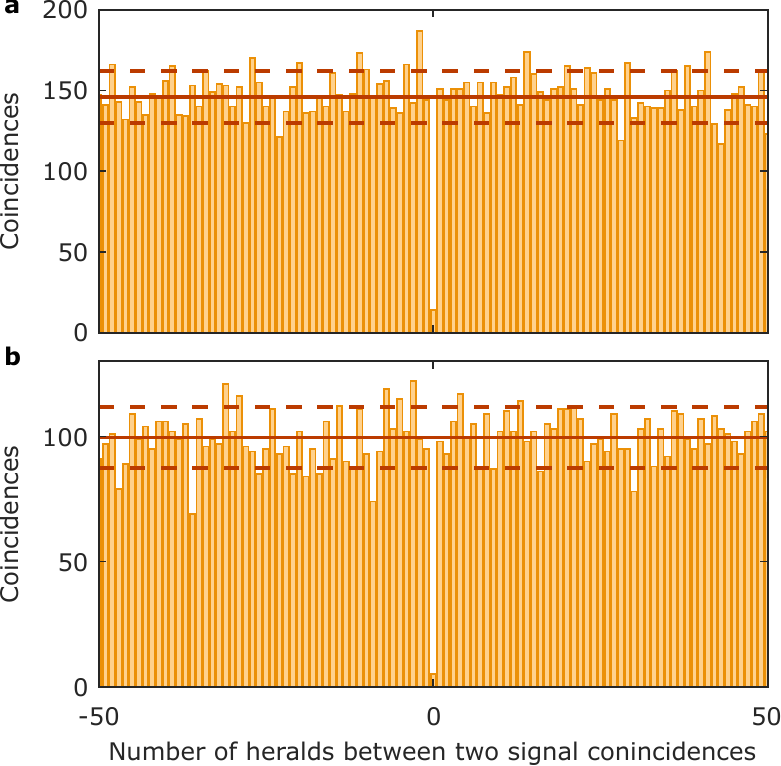}
	\caption{Heralded auto-correlation. The heralded autocorrelation of signal photons (a) going through a transparency window in the quantum memory, or (b) stored in a 3 $\mu$s AFC. The improvement after AFC storage is a consequence of the filtering effect of the quantum memory, which shifts the interference region away from the SPDC noise. The solid line represents the average of the histograms at long delays, while the dashed lines indicate one standard deviation variation from the average.}
	\label{fig:her}
\end{figure}
We started by measuring the efficiency of the AFC at different storage times, and up to 28~$\mu$s (see Fig.~\ref{fig:2}(a)). By fitting to the function $e^{-4t/T^\tx{eff}_2}$ \cite{Jobez2016} we calculate the effective coherence time of the AFC storage to be 27(3) $\mu$s, an improvement of a factor of 4 with respect to our previous work in the same waveguide without fibre-coupling \cite{Seri2018}, which prevented the use of more complex vibration isolation systems.
The maximum measured internal efficiency was 18\%, corresponding to an end-to-end efficiency of 3.6\%\ \cite{SuppMat}. We also measured the second-order cross-correlation function \gafc\ between idler and retrieved AFC echo. The results are reported in Fig.~\ref{fig:2}(b), and show that even for our longest storage time we maintained non-classical correlations, i.e. \gafc$ > 2$.
Furthermore, we measured the heralded auto-correlation function of the signal photons \cite{Fasel2004}, shown in Fig.~\ref{fig:her}. We measured values of 0.10(3) and 0.05(2) for the case of signal photons going through a transparency window and after storage for 3 $\mu$s, respectively. The improvement after the storage is due to the filtering effect of the quantum memory, which shifts the echo away from the SPDC noise.

We then carried out a full tomography of the entangled state of telecom idler and stored signal photon. We used the Franson scheme \cite{Franson1989} for the analysis, which involves the use of two unbalanced Mach-Zehnder interferometers, one each for the idler and signal photons. The detection of a photon after each interferometer post-selects the energy-time entangled state to:
\begin{equation}
\label{eq:psi}
    \ket{\Phi^+} = \frac{\ket{e_i e_s} + \ket{l_i l_s}}{\sqrt{2}}
\end{equation}
where $\ket{e}$ (early) and $\ket{l}$ (late) are the two time-bins associated with the short and long arm of the interferometer. The state $\ket{\Phi^+}$ then corresponds to a case where the signal and idler photons are in the same time bin, either $\ket{e}$ or $\ket{l}$, with equal probability. Our interferometers are fibre-based \cite{Rakonjac2021a}, with the long arm introducing a delay of 420 ns. The relative phase between the two arms can be controlled using piezo-electric fibre stretchers, which changes the base in which idler and signal modes are measured.

\begin{figure*}
	\centering
	\includegraphics[width=2\columnwidth]{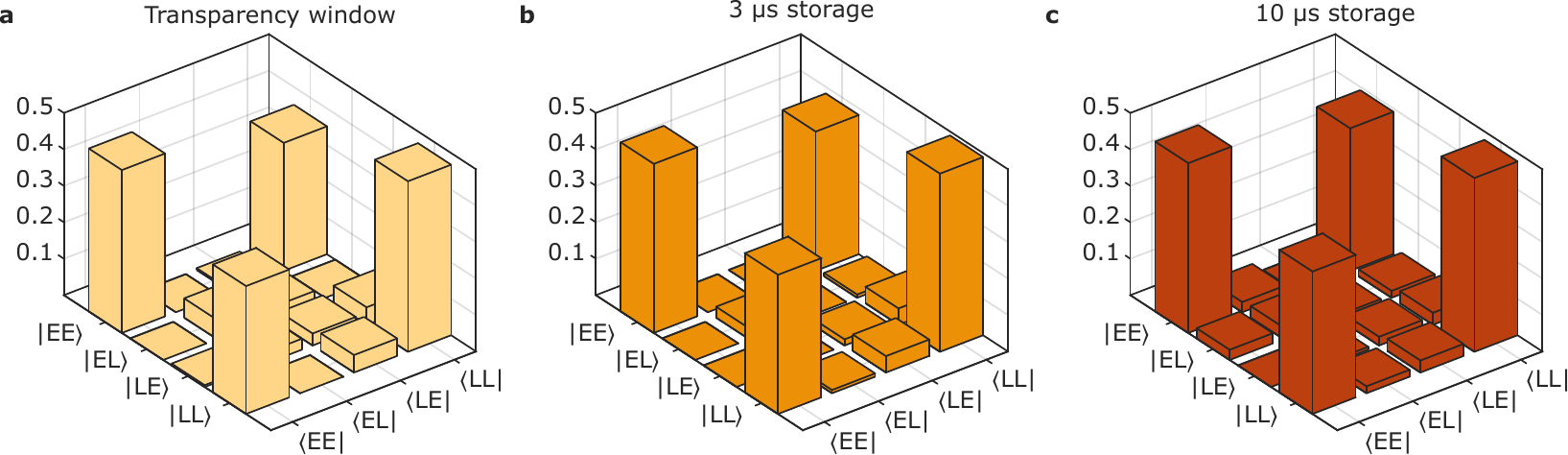}
	\caption{Entangled state tomography. (a) Reconstructed density matrix of the input state, where the signal photon passes through a transparency window in the memory crystal. The data was acquired during 30 min. (b)-(c) Density matrices for the retrieved AFC echo for storage times of 3 $\mu$s and 10 $\mu$s, respectively. Data acquisition was completed in 120 and 130 min, respectively. Only the real part of the matrices are reported.}
	\label{fig:3}
\end{figure*}

The results of the tomography are reported in Fig.~\ref{fig:3}. We started by characterising the heralded qubit and the entangled state generated by the source by preparing a transparency window in the \PrYSO\ crystal. We measured an average 1-qubit fidelity for the heralded signal qubit of $\mathcal{F}^\tx{ 1Q}_\tx{in} = 87.8(7)\%$, and a 2-qubit fidelity to the ideal state of eq. \eqref{eq:psi} of $\mathcal{F}^\tx{ 2Q}_\tx{in} = 75(2)\%$ (see \cite{SuppMat} for further details). These values are obtained from raw data and are affected by technical limitations in our analysers, especially the interferometer for the signal mode, as there is a ratio of 50\% between the transmission of the long and short arm that limits the visibility to 95\%. If we correct for these imperfections and for the dark counts of the detectors \cite{SuppMat}, the fidelities become $\widetilde{\mathcal{F}}^\tx{ 1Q}_\tx{in} = 91.1(7)\%$ and $\widetilde{\mathcal{F}}^\tx{ 2Q}_\tx{in} = 81(2)\%$. These values are still limited by the broadband noise generated by the SPDC, and by the limited coherence of the SPDC pump laser.

We then prepared an AFC  in the quantum memory with 3 $\mu$s of storage time and an efficiency of 13.7(4)\%. We repeated the tomography, and measured raw fidelities of $\mathcal{F}^\tx{ 1Q}_\tx{ 3$\mu$s} = 89.7(6)\%$ and $\mathcal{F}^\tx{ 2Q}_\tx{ 3$\mu$s} = 79(2)\%$. If corrected, we obtain the values $\widetilde{\mathcal{F}}^\tx{ 1Q}_\tx{3$\mu$s} = 93.8(7)\%$ and $\widetilde{\mathcal{F}}^\tx{ 2Q}_\tx{3$\mu$s} = 86(2)\%$. These are still limited by the \gafc\ of the retrieved echo after the analyser, and by the quality of the entangled state preparation. However the fidelities are improved  due to the aforementioned temporal filtering of the AFC. Finally, we extended the storage time of the AFC to 10 $\mu$s, with an efficiency of 4.2(2)\%, and repeated the tomography. We obtained fidelities of $\mathcal{F}^\tx{ 1Q}_\tx{10$\mu$s} = 88(2)\%$ and $\mathcal{F}^\tx{ 2Q}_\tx{10$\mu$s} = 77(3)\%$ for the raw data, and of $\widetilde{\mathcal{F}}^\tx{ 1Q}_\tx{10$\mu$s} = 94(2)\%$ and $\widetilde{\mathcal{F}}^\tx{ 2Q}_\tx{10$\mu$s} = 86(4)\%$ for the corrected case.

The AFC temporal filtering complicates the comparison of the entangled state before and after the storage, as the input-output fidelity would be lowered, despite an improved fidelity with the ideal state after storage. Subtracting the SPDC background from the input state provides a better comparison for the temporally-filtered stored AFC state. In this case, the input/output 2-qubit fidelity $\widetilde{\mathcal{F}}^\tx{ 2Q}_\tx{i/o}$ is 98\% and 97\% for the 3 and 10 $\mu$s storage times, respectively, confirming the reliability of the AFC process and of our device.

\section{Discussion}
In this work, we have demonstrated a fibre-integrated quantum memory based on laser-written waveguides. Our results confirm that the entanglement generated by an SPDC process was maintained after a predetermined storage of up to 10 $\mu$s in the fibre-coupled integrated memory. We demonstrated an increase in storage time of three orders of magnitude with respect to previous realizations of fibre-integrated quantum memories and an improvement of at least 20 times in terms of efficiency \cite{Saglamyurek2011, Saglamyurek2015}, with comparable values of fidelities.

A natural step forward would be to perform on-demand read-out of the stored excitation through storage in a spin ground state of Pr \cite{Gundogan2015,Seri2017,Rakonjac2021a}. This has only been demonstrated in a few integrated systems: in a type I \cite{Seri2019thesis} and type II waveguide \cite{Corrielli2016} in \PrYSO\ and in a type IV waveguide in $^{151}$Eu$^{3+}$:Y$_2$SiO$_5$\cite{Liu2020b}, however these works were limited to the storage of classical light. The noise generated by the strong control pulses required for spin-wave storage poses a particular challenge to integrated devices, since the stored single photons and the bright control pulses are in the same spatial mode. On the other hand, the integrated nature of the memory could facilitate the inclusion of narrow laser-written filters, directional couplers or electrodes for noise-free on-demand storage \cite{Liu2020a, Liu2020b}.

In conclusion, our approach provides an excellent device, merging the well-explored world of laser-written components with the excellent properties of rare-earth based memories and the reliability of fibre-integrated devices. This combination could allow for an extreme level of multiplexing: given their size, hundreds of waveguides could be written in a single crystal, exploiting the three-dimensional nature of the fabrication process. This would further maximize the exploitability of temporal and frequency modes \cite{Sinclair2014, Seri2019}. An integrated and fibre-coupled quantum memory could be directly interfaced with several fibre-based components, such as fibre arrays, or superconducting detectors, which could be included in the same cryostat without the need for free-space optical access. Moreover, type I waveguides feature lower bending losses with respect to other types \cite{Corrielli2016, Seri2018, Liu2020a}, which make our device compatible with more involved optical circuitry, allowing for signal routing and filtering. Finally, spin-wave storage in Pr opens the way to ultra-long storage times \cite{Laplane2017, Ma2021, Ortu2021}. Our system will then encompass all the basic requirements for a quantum repeater building-block, leveraging on its multimodality, fibre-compatibility and scalability.

\section{Acknowledgments}
This project received funding from the European Research Council (ERC) via the Advanced Grant CAPABLE (742745), from the European Union Horizon 2020 research and innovation programme within the Flagship on Quantum Technologies through grant 820445 (QIA), under grant agreement no. 899275 (DAALI) of the FET-Open funding call, and under the Marie Sk\l odowska-Curie grant agreement no. 713729 (ICFOStepstone 2) and no. 758461 (proBIST), from Laserlab Europe (grant agreement no. 654148), from the Gordon and Betty Moore Foundation through grant GBMF7446 to H.d.R., and from the Government of Spain (PID2019-106850RB-I00, Severo Ochoa CEX2019-000910-S, BES-2017-082464), Fundaci\'o Cellex, Fundaci\'o Mir-Puig, and Generalitat de Catalunya (CERCA, AGAUR, Quantum CAT).


%


\begin{thebibliography}{41}%
\makeatletter
\providecommand \@ifxundefined [1]{%
 \@ifx{#1\undefined}
}%
\providecommand \@ifnum [1]{%
 \ifnum #1\expandafter \@firstoftwo
 \else \expandafter \@secondoftwo
 \fi
}%
\providecommand \@ifx [1]{%
 \ifx #1\expandafter \@firstoftwo
 \else \expandafter \@secondoftwo
 \fi
}%
\providecommand \natexlab [1]{#1}%
\providecommand \enquote  [1]{``#1''}%
\providecommand \bibnamefont  [1]{#1}%
\providecommand \bibfnamefont [1]{#1}%
\providecommand \citenamefont [1]{#1}%
\providecommand \href@noop [0]{\@secondoftwo}%
\providecommand \href [0]{\begingroup \@sanitize@url \@href}%
\providecommand \@href[1]{\@@startlink{#1}\@@href}%
\providecommand \@@href[1]{\endgroup#1\@@endlink}%
\providecommand \@sanitize@url [0]{\catcode `\\12\catcode `\$12\catcode
  `\&12\catcode `\#12\catcode `\^12\catcode `\_12\catcode `\%12\relax}%
\providecommand \@@startlink[1]{}%
\providecommand \@@endlink[0]{}%
\providecommand \url  [0]{\begingroup\@sanitize@url \@url }%
\providecommand \@url [1]{\endgroup\@href {#1}{\urlprefix }}%
\providecommand \urlprefix  [0]{URL }%
\providecommand \Eprint [0]{\href }%
\providecommand \doibase [0]{https://doi.org/}%
\providecommand \selectlanguage [0]{\@gobble}%
\providecommand \bibinfo  [0]{\@secondoftwo}%
\providecommand \bibfield  [0]{\@secondoftwo}%
\providecommand \translation [1]{[#1]}%
\providecommand \BibitemOpen [0]{}%
\providecommand \bibitemStop [0]{}%
\providecommand \bibitemNoStop [0]{.\EOS\space}%
\providecommand \EOS [0]{\spacefactor3000\relax}%
\providecommand \BibitemShut  [1]{\csname bibitem#1\endcsname}%
\let\auto@bib@innerbib\@empty
\bibitem [{\citenamefont {Kimble}(2008)}]{Kimble2008}%
  \BibitemOpen
  \bibfield  {author} {\bibinfo {author} {\bibfnamefont {H.~J.}\ \bibnamefont
  {Kimble}},\ }\href {https://doi.org/10.1038/nature07127} {\bibfield
  {journal} {\bibinfo  {journal} {Nature}\ }\textbf {\bibinfo {volume} {453}},\
  \bibinfo {pages} {1023} (\bibinfo {year} {2008})}\BibitemShut {NoStop}%
\bibitem [{\citenamefont {Wehner}\ \emph {et~al.}(2018)\citenamefont {Wehner},
  \citenamefont {Elkouss},\ and\ \citenamefont {Hanson}}]{Wehner2018}%
  \BibitemOpen
  \bibfield  {author} {\bibinfo {author} {\bibfnamefont {S.}~\bibnamefont
  {Wehner}}, \bibinfo {author} {\bibfnamefont {D.}~\bibnamefont {Elkouss}},\
  and\ \bibinfo {author} {\bibfnamefont {R.}~\bibnamefont {Hanson}},\ }\href
  {https://doi.org/10.1126/science.aam9288} {\bibfield  {journal} {\bibinfo
  {journal} {Science}\ }\textbf {\bibinfo {volume} {362}},\ \bibinfo {pages}
  {eaam9288} (\bibinfo {year} {2018})}\BibitemShut {NoStop}%
\bibitem [{\citenamefont {Duan}\ \emph {et~al.}(2001)\citenamefont {Duan},
  \citenamefont {Lukin}, \citenamefont {Cirac},\ and\ \citenamefont
  {Zoller}}]{Duan2001}%
  \BibitemOpen
  \bibfield  {author} {\bibinfo {author} {\bibfnamefont {L.-M.}\ \bibnamefont
  {Duan}}, \bibinfo {author} {\bibfnamefont {M.~D.}\ \bibnamefont {Lukin}},
  \bibinfo {author} {\bibfnamefont {J.~I.}\ \bibnamefont {Cirac}},\ and\
  \bibinfo {author} {\bibfnamefont {P.}~\bibnamefont {Zoller}},\ }\href
  {https://doi.org/10.1038/35106500} {\bibfield  {journal} {\bibinfo  {journal}
  {Nature}\ }\textbf {\bibinfo {volume} {414}},\ \bibinfo {pages} {413}
  (\bibinfo {year} {2001})}\BibitemShut {NoStop}%
\bibitem [{\citenamefont {Simon}\ \emph {et~al.}(2007)\citenamefont {Simon},
  \citenamefont {de~Riedmatten}, \citenamefont {Afzelius}, \citenamefont
  {Sangouard}, \citenamefont {Zbinden},\ and\ \citenamefont
  {Gisin}}]{Simon2007}%
  \BibitemOpen
  \bibfield  {author} {\bibinfo {author} {\bibfnamefont {C.}~\bibnamefont
  {Simon}}, \bibinfo {author} {\bibfnamefont {H.}~\bibnamefont
  {de~Riedmatten}}, \bibinfo {author} {\bibfnamefont {M.}~\bibnamefont
  {Afzelius}}, \bibinfo {author} {\bibfnamefont {N.}~\bibnamefont {Sangouard}},
  \bibinfo {author} {\bibfnamefont {H.}~\bibnamefont {Zbinden}},\ and\ \bibinfo
  {author} {\bibfnamefont {N.}~\bibnamefont {Gisin}},\ }\href
  {https://doi.org/10.1103/PhysRevLett.98.190503} {\bibfield  {journal}
  {\bibinfo  {journal} {Physical Review Letters}\ }\textbf {\bibinfo {volume}
  {98}},\ \bibinfo {pages} {190503} (\bibinfo {year} {2007})}\BibitemShut
  {NoStop}%
\bibitem [{\citenamefont {Sangouard}\ \emph {et~al.}(2011)\citenamefont
  {Sangouard}, \citenamefont {Simon}, \citenamefont {de~Riedmatten},\ and\
  \citenamefont {Gisin}}]{Sangouard2011}%
  \BibitemOpen
  \bibfield  {author} {\bibinfo {author} {\bibfnamefont {N.}~\bibnamefont
  {Sangouard}}, \bibinfo {author} {\bibfnamefont {C.}~\bibnamefont {Simon}},
  \bibinfo {author} {\bibfnamefont {H.}~\bibnamefont {de~Riedmatten}},\ and\
  \bibinfo {author} {\bibfnamefont {N.}~\bibnamefont {Gisin}},\ }\href
  {http://link.aps.org/doi/10.1103/RevModPhys.83.33} {\bibfield  {journal}
  {\bibinfo  {journal} {Rev. Mod. Phys.}\ }\textbf {\bibinfo {volume} {83}},\
  \bibinfo {pages} {33} (\bibinfo {year} {2011})}\BibitemShut {NoStop}%
\bibitem [{\citenamefont {Corzo}\ \emph {et~al.}(2019)\citenamefont {Corzo},
  \citenamefont {Raskop}, \citenamefont {Chandra}, \citenamefont {Sheremet},
  \citenamefont {Gouraud},\ and\ \citenamefont {Laurat}}]{Corzo2019}%
  \BibitemOpen
  \bibfield  {author} {\bibinfo {author} {\bibfnamefont {N.~V.}\ \bibnamefont
  {Corzo}}, \bibinfo {author} {\bibfnamefont {J.}~\bibnamefont {Raskop}},
  \bibinfo {author} {\bibfnamefont {A.}~\bibnamefont {Chandra}}, \bibinfo
  {author} {\bibfnamefont {A.~S.}\ \bibnamefont {Sheremet}}, \bibinfo {author}
  {\bibfnamefont {B.}~\bibnamefont {Gouraud}},\ and\ \bibinfo {author}
  {\bibfnamefont {J.}~\bibnamefont {Laurat}},\ }\href
  {https://doi.org/10.1038/s41586-019-0902-3} {\bibfield  {journal} {\bibinfo
  {journal} {Nature}\ }\textbf {\bibinfo {volume} {566}},\ \bibinfo {pages}
  {359} (\bibinfo {year} {2019})}\BibitemShut {NoStop}%
\bibitem [{\citenamefont {Uppu}\ \emph {et~al.}(2020)\citenamefont {Uppu},
  \citenamefont {Pedersen}, \citenamefont {Wang}, \citenamefont {Olesen},
  \citenamefont {Papon}, \citenamefont {Zhou}, \citenamefont {Midolo},
  \citenamefont {Scholz}, \citenamefont {Wieck}, \citenamefont {Ludwig},\ and\
  \citenamefont {Lodahl}}]{Uppu2020}%
  \BibitemOpen
  \bibfield  {author} {\bibinfo {author} {\bibfnamefont {R.}~\bibnamefont
  {Uppu}}, \bibinfo {author} {\bibfnamefont {F.~T.}\ \bibnamefont {Pedersen}},
  \bibinfo {author} {\bibfnamefont {Y.}~\bibnamefont {Wang}}, \bibinfo {author}
  {\bibfnamefont {C.~T.}\ \bibnamefont {Olesen}}, \bibinfo {author}
  {\bibfnamefont {C.}~\bibnamefont {Papon}}, \bibinfo {author} {\bibfnamefont
  {X.}~\bibnamefont {Zhou}}, \bibinfo {author} {\bibfnamefont {L.}~\bibnamefont
  {Midolo}}, \bibinfo {author} {\bibfnamefont {S.}~\bibnamefont {Scholz}},
  \bibinfo {author} {\bibfnamefont {A.~D.}\ \bibnamefont {Wieck}}, \bibinfo
  {author} {\bibfnamefont {A.}~\bibnamefont {Ludwig}},\ and\ \bibinfo {author}
  {\bibfnamefont {P.}~\bibnamefont {Lodahl}},\ }\href
  {https://doi.org/10.1126/sciadv.abc8268} {\bibfield  {journal} {\bibinfo
  {journal} {Science Advances}\ }\textbf {\bibinfo {volume} {6}},\ \bibinfo
  {pages} {eabc8268} (\bibinfo {year} {2020})},\ \Eprint
  {https://arxiv.org/abs/2003.08919} {arXiv:2003.08919} \BibitemShut {NoStop}%
\bibitem [{\citenamefont {You}\ \emph {et~al.}(2021)\citenamefont {You},
  \citenamefont {Zheng}, \citenamefont {Chen}, \citenamefont {Liu},
  \citenamefont {Qin}, \citenamefont {Xu}, \citenamefont {Ge}, \citenamefont
  {Chung}, \citenamefont {Qiao}, \citenamefont {Jiang}, \citenamefont {Zhong},
  \citenamefont {Chen}, \citenamefont {Wang}, \citenamefont {He}, \citenamefont
  {Xie}, \citenamefont {Li}, \citenamefont {You}, \citenamefont {Schneider},
  \citenamefont {Yin}, \citenamefont {Chen}, \citenamefont {Benyoucef},
  \citenamefont {Huo}, \citenamefont {Hoefling}, \citenamefont {Zhang},
  \citenamefont {Lu},\ and\ \citenamefont {Pan}}]{You2021}%
  \BibitemOpen
  \bibfield  {author} {\bibinfo {author} {\bibfnamefont {X.}~\bibnamefont
  {You}}, \bibinfo {author} {\bibfnamefont {M.-Y.}\ \bibnamefont {Zheng}},
  \bibinfo {author} {\bibfnamefont {S.}~\bibnamefont {Chen}}, \bibinfo {author}
  {\bibfnamefont {R.-Z.}\ \bibnamefont {Liu}}, \bibinfo {author} {\bibfnamefont
  {J.}~\bibnamefont {Qin}}, \bibinfo {author} {\bibfnamefont {M.~C.}\
  \bibnamefont {Xu}}, \bibinfo {author} {\bibfnamefont {Z.~X.}\ \bibnamefont
  {Ge}}, \bibinfo {author} {\bibfnamefont {T.~H.}\ \bibnamefont {Chung}},
  \bibinfo {author} {\bibfnamefont {Y.~K.}\ \bibnamefont {Qiao}}, \bibinfo
  {author} {\bibfnamefont {Y.~F.}\ \bibnamefont {Jiang}}, \bibinfo {author}
  {\bibfnamefont {H.~S.}\ \bibnamefont {Zhong}}, \bibinfo {author}
  {\bibfnamefont {M.~C.}\ \bibnamefont {Chen}}, \bibinfo {author}
  {\bibfnamefont {H.}~\bibnamefont {Wang}}, \bibinfo {author} {\bibfnamefont
  {Y.~M.}\ \bibnamefont {He}}, \bibinfo {author} {\bibfnamefont {X.~P.}\
  \bibnamefont {Xie}}, \bibinfo {author} {\bibfnamefont {H.}~\bibnamefont
  {Li}}, \bibinfo {author} {\bibfnamefont {L.~X.}\ \bibnamefont {You}},
  \bibinfo {author} {\bibfnamefont {C.}~\bibnamefont {Schneider}}, \bibinfo
  {author} {\bibfnamefont {J.}~\bibnamefont {Yin}}, \bibinfo {author}
  {\bibfnamefont {T.~Y.}\ \bibnamefont {Chen}}, \bibinfo {author}
  {\bibfnamefont {M.}~\bibnamefont {Benyoucef}}, \bibinfo {author}
  {\bibfnamefont {Y.-H.}\ \bibnamefont {Huo}}, \bibinfo {author} {\bibfnamefont
  {S.}~\bibnamefont {Hoefling}}, \bibinfo {author} {\bibfnamefont
  {Q.}~\bibnamefont {Zhang}}, \bibinfo {author} {\bibfnamefont {C.-Y.}\
  \bibnamefont {Lu}},\ and\ \bibinfo {author} {\bibfnamefont {J.-W.}\
  \bibnamefont {Pan}},\ }\href {http://arxiv.org/abs/2106.15545} {\bibfield
  {journal} {\bibinfo  {journal} {arxiv:2106.15545}\ } (\bibinfo {year}
  {2021})}\BibitemShut {NoStop}%
\bibitem [{\citenamefont {Sipahigil}\ \emph {et~al.}(2016)\citenamefont
  {Sipahigil}, \citenamefont {Evans}, \citenamefont {Sukachev}, \citenamefont
  {Burek}, \citenamefont {Borregaard}, \citenamefont {Bhaskar}, \citenamefont
  {Nguyen}, \citenamefont {Pacheco}, \citenamefont {Atikian}, \citenamefont
  {Meuwly}, \citenamefont {Camacho}, \citenamefont {Jelezko}, \citenamefont
  {Bielejec}, \citenamefont {Park}, \citenamefont {Lon{\v{c}}ar},\ and\
  \citenamefont {Lukin}}]{Sipahigil2016}%
  \BibitemOpen
  \bibfield  {author} {\bibinfo {author} {\bibfnamefont {A.}~\bibnamefont
  {Sipahigil}}, \bibinfo {author} {\bibfnamefont {R.~E.}\ \bibnamefont
  {Evans}}, \bibinfo {author} {\bibfnamefont {D.~D.}\ \bibnamefont {Sukachev}},
  \bibinfo {author} {\bibfnamefont {M.~J.}\ \bibnamefont {Burek}}, \bibinfo
  {author} {\bibfnamefont {J.}~\bibnamefont {Borregaard}}, \bibinfo {author}
  {\bibfnamefont {M.~K.}\ \bibnamefont {Bhaskar}}, \bibinfo {author}
  {\bibfnamefont {C.~T.}\ \bibnamefont {Nguyen}}, \bibinfo {author}
  {\bibfnamefont {J.~L.}\ \bibnamefont {Pacheco}}, \bibinfo {author}
  {\bibfnamefont {H.~A.}\ \bibnamefont {Atikian}}, \bibinfo {author}
  {\bibfnamefont {C.}~\bibnamefont {Meuwly}}, \bibinfo {author} {\bibfnamefont
  {R.~M.}\ \bibnamefont {Camacho}}, \bibinfo {author} {\bibfnamefont
  {F.}~\bibnamefont {Jelezko}}, \bibinfo {author} {\bibfnamefont
  {E.}~\bibnamefont {Bielejec}}, \bibinfo {author} {\bibfnamefont
  {H.}~\bibnamefont {Park}}, \bibinfo {author} {\bibfnamefont {M.}~\bibnamefont
  {Lon{\v{c}}ar}},\ and\ \bibinfo {author} {\bibfnamefont {M.~D.}\ \bibnamefont
  {Lukin}},\ }\href {https://doi.org/10.1126/science.aah6875} {\bibfield
  {journal} {\bibinfo  {journal} {Science}\ }\textbf {\bibinfo {volume}
  {354}},\ \bibinfo {pages} {847} (\bibinfo {year} {2016})}\BibitemShut
  {NoStop}%
\bibitem [{\citenamefont {Bhaskar}\ \emph {et~al.}(2020)\citenamefont
  {Bhaskar}, \citenamefont {Riedinger}, \citenamefont {Machielse},
  \citenamefont {Levonian}, \citenamefont {Nguyen}, \citenamefont {Knall},
  \citenamefont {Park}, \citenamefont {Englund}, \citenamefont {Lon{\v{c}}ar},
  \citenamefont {Sukachev},\ and\ \citenamefont {Lukin}}]{Bhaskar2020}%
  \BibitemOpen
  \bibfield  {author} {\bibinfo {author} {\bibfnamefont {M.~K.}\ \bibnamefont
  {Bhaskar}}, \bibinfo {author} {\bibfnamefont {R.}~\bibnamefont {Riedinger}},
  \bibinfo {author} {\bibfnamefont {B.}~\bibnamefont {Machielse}}, \bibinfo
  {author} {\bibfnamefont {D.~S.}\ \bibnamefont {Levonian}}, \bibinfo {author}
  {\bibfnamefont {C.~T.}\ \bibnamefont {Nguyen}}, \bibinfo {author}
  {\bibfnamefont {E.~N.}\ \bibnamefont {Knall}}, \bibinfo {author}
  {\bibfnamefont {H.}~\bibnamefont {Park}}, \bibinfo {author} {\bibfnamefont
  {D.}~\bibnamefont {Englund}}, \bibinfo {author} {\bibfnamefont
  {M.}~\bibnamefont {Lon{\v{c}}ar}}, \bibinfo {author} {\bibfnamefont {D.~D.}\
  \bibnamefont {Sukachev}},\ and\ \bibinfo {author} {\bibfnamefont {M.~D.}\
  \bibnamefont {Lukin}},\ }\href {https://doi.org/10.1038/s41586-020-2103-5}
  {\bibfield  {journal} {\bibinfo  {journal} {Nature}\ }\textbf {\bibinfo
  {volume} {580}},\ \bibinfo {pages} {60} (\bibinfo {year} {2020})}\BibitemShut
  {NoStop}%
\bibitem [{\citenamefont {Macfarlane}(2002)}]{Macfarlane2002}%
  \BibitemOpen
  \bibfield  {author} {\bibinfo {author} {\bibfnamefont {R.~M.}\ \bibnamefont
  {Macfarlane}},\ }\href
  {http://www.sciencedirect.com/science/article/B6TJH-479STRG-3/2/1570a648518b1c4a5e2cb36499fc00ae}
  {\bibfield  {journal} {\bibinfo  {journal} {J. Lumin.}\ }\textbf {\bibinfo
  {volume} {100}},\ \bibinfo {pages} {1} (\bibinfo {year} {2002})}\BibitemShut
  {NoStop}%
\bibitem [{\citenamefont {Clausen}\ \emph {et~al.}(2011)\citenamefont
  {Clausen}, \citenamefont {Usmani}, \citenamefont {Bussi{\`{e}}res},
  \citenamefont {Sangouard}, \citenamefont {Afzelius}, \citenamefont
  {de~Riedmatten},\ and\ \citenamefont {Gisin}}]{Clausen2011}%
  \BibitemOpen
  \bibfield  {author} {\bibinfo {author} {\bibfnamefont {C.}~\bibnamefont
  {Clausen}}, \bibinfo {author} {\bibfnamefont {I.}~\bibnamefont {Usmani}},
  \bibinfo {author} {\bibfnamefont {F.}~\bibnamefont {Bussi{\`{e}}res}},
  \bibinfo {author} {\bibfnamefont {N.}~\bibnamefont {Sangouard}}, \bibinfo
  {author} {\bibfnamefont {M.}~\bibnamefont {Afzelius}}, \bibinfo {author}
  {\bibfnamefont {H.}~\bibnamefont {de~Riedmatten}},\ and\ \bibinfo {author}
  {\bibfnamefont {N.}~\bibnamefont {Gisin}},\ }\href
  {https://doi.org/10.1038/nature09662} {\bibfield  {journal} {\bibinfo
  {journal} {Nature}\ }\textbf {\bibinfo {volume} {469}},\ \bibinfo {pages}
  {508} (\bibinfo {year} {2011})}\BibitemShut {NoStop}%
\bibitem [{\citenamefont {Usmani}\ \emph {et~al.}(2012)\citenamefont {Usmani},
  \citenamefont {Clausen}, \citenamefont {Bussi{\`{e}}res}, \citenamefont
  {Sangouard}, \citenamefont {Afzelius},\ and\ \citenamefont
  {Gisin}}]{Usmani2012}%
  \BibitemOpen
  \bibfield  {author} {\bibinfo {author} {\bibfnamefont {I.}~\bibnamefont
  {Usmani}}, \bibinfo {author} {\bibfnamefont {C.}~\bibnamefont {Clausen}},
  \bibinfo {author} {\bibfnamefont {F.}~\bibnamefont {Bussi{\`{e}}res}},
  \bibinfo {author} {\bibfnamefont {N.}~\bibnamefont {Sangouard}}, \bibinfo
  {author} {\bibfnamefont {M.}~\bibnamefont {Afzelius}},\ and\ \bibinfo
  {author} {\bibfnamefont {N.}~\bibnamefont {Gisin}},\ }\href
  {https://doi.org/10.1038/nphoton.2012.34} {\bibfield  {journal} {\bibinfo
  {journal} {Nature Photonics}\ }\textbf {\bibinfo {volume} {6}},\ \bibinfo
  {pages} {234} (\bibinfo {year} {2012})}\BibitemShut {NoStop}%
\bibitem [{\citenamefont {Kutluer}\ \emph {et~al.}(2019)\citenamefont
  {Kutluer}, \citenamefont {Distante}, \citenamefont {Casabone}, \citenamefont
  {Duranti}, \citenamefont {Mazzera},\ and\ \citenamefont
  {de~Riedmatten}}]{Kutluer2019}%
  \BibitemOpen
  \bibfield  {author} {\bibinfo {author} {\bibfnamefont {K.}~\bibnamefont
  {Kutluer}}, \bibinfo {author} {\bibfnamefont {E.}~\bibnamefont {Distante}},
  \bibinfo {author} {\bibfnamefont {B.}~\bibnamefont {Casabone}}, \bibinfo
  {author} {\bibfnamefont {S.}~\bibnamefont {Duranti}}, \bibinfo {author}
  {\bibfnamefont {M.}~\bibnamefont {Mazzera}},\ and\ \bibinfo {author}
  {\bibfnamefont {H.}~\bibnamefont {de~Riedmatten}},\ }\href
  {https://doi.org/10.1103/PhysRevLett.123.030501} {\bibfield  {journal}
  {\bibinfo  {journal} {Physical Review Letters}\ }\textbf {\bibinfo {volume}
  {123}},\ \bibinfo {pages} {030501} (\bibinfo {year} {2019})}\BibitemShut
  {NoStop}%
\bibitem [{\citenamefont {Lago-Rivera}\ \emph {et~al.}(2021)\citenamefont
  {Lago-Rivera}, \citenamefont {Grandi}, \citenamefont {Rakonjac},
  \citenamefont {Seri},\ and\ \citenamefont {de~Riedmatten}}]{Lago-Rivera2021}%
  \BibitemOpen
  \bibfield  {author} {\bibinfo {author} {\bibfnamefont {D.}~\bibnamefont
  {Lago-Rivera}}, \bibinfo {author} {\bibfnamefont {S.}~\bibnamefont {Grandi}},
  \bibinfo {author} {\bibfnamefont {J.~V.}\ \bibnamefont {Rakonjac}}, \bibinfo
  {author} {\bibfnamefont {A.}~\bibnamefont {Seri}},\ and\ \bibinfo {author}
  {\bibfnamefont {H.}~\bibnamefont {de~Riedmatten}},\ }\href
  {https://doi.org/10.1038/s41586-021-03481-8} {\bibfield  {journal} {\bibinfo
  {journal} {Nature}\ }\textbf {\bibinfo {volume} {594}},\ \bibinfo {pages}
  {37} (\bibinfo {year} {2021})}\BibitemShut {NoStop}%
\bibitem [{\citenamefont {Liu}\ \emph {et~al.}(2021)\citenamefont {Liu},
  \citenamefont {Hu}, \citenamefont {Li}, \citenamefont {Li}, \citenamefont
  {Li}, \citenamefont {Liang}, \citenamefont {Zhou}, \citenamefont {Li},\ and\
  \citenamefont {Guo}}]{Liu2021}%
  \BibitemOpen
  \bibfield  {author} {\bibinfo {author} {\bibfnamefont {X.}~\bibnamefont
  {Liu}}, \bibinfo {author} {\bibfnamefont {J.}~\bibnamefont {Hu}}, \bibinfo
  {author} {\bibfnamefont {Z.-F.}\ \bibnamefont {Li}}, \bibinfo {author}
  {\bibfnamefont {X.}~\bibnamefont {Li}}, \bibinfo {author} {\bibfnamefont
  {P.-Y.}\ \bibnamefont {Li}}, \bibinfo {author} {\bibfnamefont {P.-J.}\
  \bibnamefont {Liang}}, \bibinfo {author} {\bibfnamefont {Z.-Q.}\ \bibnamefont
  {Zhou}}, \bibinfo {author} {\bibfnamefont {C.-F.}\ \bibnamefont {Li}},\ and\
  \bibinfo {author} {\bibfnamefont {G.-C.}\ \bibnamefont {Guo}},\ }\href
  {https://doi.org/10.1038/s41586-021-03505-3} {\bibfield  {journal} {\bibinfo
  {journal} {Nature}\ }\textbf {\bibinfo {volume} {594}},\ \bibinfo {pages}
  {41} (\bibinfo {year} {2021})}\BibitemShut {NoStop}%
\bibitem [{\citenamefont {Seri}\ \emph {et~al.}(2017)\citenamefont {Seri},
  \citenamefont {Lenhard}, \citenamefont {Riel{\"{a}}nder}, \citenamefont
  {G{\"{u}}ndoğan}, \citenamefont {Ledingham}, \citenamefont {Mazzera},\ and\
  \citenamefont {de~Riedmatten}}]{Seri2017}%
  \BibitemOpen
  \bibfield  {author} {\bibinfo {author} {\bibfnamefont {A.}~\bibnamefont
  {Seri}}, \bibinfo {author} {\bibfnamefont {A.}~\bibnamefont {Lenhard}},
  \bibinfo {author} {\bibfnamefont {D.}~\bibnamefont {Riel{\"{a}}nder}},
  \bibinfo {author} {\bibfnamefont {M.}~\bibnamefont {G{\"{u}}ndoğan}},
  \bibinfo {author} {\bibfnamefont {P.~M.}\ \bibnamefont {Ledingham}}, \bibinfo
  {author} {\bibfnamefont {M.}~\bibnamefont {Mazzera}},\ and\ \bibinfo {author}
  {\bibfnamefont {H.}~\bibnamefont {de~Riedmatten}},\ }\href
  {https://doi.org/10.1103/PhysRevX.7.021028} {\bibfield  {journal} {\bibinfo
  {journal} {Physical Review X}\ }\textbf {\bibinfo {volume} {7}},\ \bibinfo
  {pages} {021028} (\bibinfo {year} {2017})}\BibitemShut {NoStop}%
\bibitem [{\citenamefont {Yang}\ \emph {et~al.}(2018)\citenamefont {Yang},
  \citenamefont {Zhou}, \citenamefont {Hua}, \citenamefont {Liu}, \citenamefont
  {Li}, \citenamefont {Li}, \citenamefont {Ma}, \citenamefont {Liu},
  \citenamefont {Liang}, \citenamefont {Li}, \citenamefont {Xiao},
  \citenamefont {Hu}, \citenamefont {Li},\ and\ \citenamefont
  {Guo}}]{Yang2018}%
  \BibitemOpen
  \bibfield  {author} {\bibinfo {author} {\bibfnamefont {T.-S.}\ \bibnamefont
  {Yang}}, \bibinfo {author} {\bibfnamefont {Z.-Q.}\ \bibnamefont {Zhou}},
  \bibinfo {author} {\bibfnamefont {Y.-L.}\ \bibnamefont {Hua}}, \bibinfo
  {author} {\bibfnamefont {X.}~\bibnamefont {Liu}}, \bibinfo {author}
  {\bibfnamefont {Z.-F.}\ \bibnamefont {Li}}, \bibinfo {author} {\bibfnamefont
  {P.-Y.}\ \bibnamefont {Li}}, \bibinfo {author} {\bibfnamefont
  {Y.}~\bibnamefont {Ma}}, \bibinfo {author} {\bibfnamefont {C.}~\bibnamefont
  {Liu}}, \bibinfo {author} {\bibfnamefont {P.-J.}\ \bibnamefont {Liang}},
  \bibinfo {author} {\bibfnamefont {X.}~\bibnamefont {Li}}, \bibinfo {author}
  {\bibfnamefont {Y.-X.}\ \bibnamefont {Xiao}}, \bibinfo {author}
  {\bibfnamefont {J.}~\bibnamefont {Hu}}, \bibinfo {author} {\bibfnamefont
  {C.-F.}\ \bibnamefont {Li}},\ and\ \bibinfo {author} {\bibfnamefont {G.-C.}\
  \bibnamefont {Guo}},\ }\href {https://doi.org/10.1038/s41467-018-05669-5}
  {\bibfield  {journal} {\bibinfo  {journal} {Nature Communications}\ }\textbf
  {\bibinfo {volume} {9}},\ \bibinfo {pages} {3407} (\bibinfo {year}
  {2018})}\BibitemShut {NoStop}%
\bibitem [{\citenamefont {Seri}\ \emph {et~al.}(2019)\citenamefont {Seri},
  \citenamefont {Lago-Rivera}, \citenamefont {Lenhard}, \citenamefont
  {Corrielli}, \citenamefont {Osellame}, \citenamefont {Mazzera},\ and\
  \citenamefont {de~Riedmatten}}]{Seri2019}%
  \BibitemOpen
  \bibfield  {author} {\bibinfo {author} {\bibfnamefont {A.}~\bibnamefont
  {Seri}}, \bibinfo {author} {\bibfnamefont {D.}~\bibnamefont {Lago-Rivera}},
  \bibinfo {author} {\bibfnamefont {A.}~\bibnamefont {Lenhard}}, \bibinfo
  {author} {\bibfnamefont {G.}~\bibnamefont {Corrielli}}, \bibinfo {author}
  {\bibfnamefont {R.}~\bibnamefont {Osellame}}, \bibinfo {author}
  {\bibfnamefont {M.}~\bibnamefont {Mazzera}},\ and\ \bibinfo {author}
  {\bibfnamefont {H.}~\bibnamefont {de~Riedmatten}},\ }\href
  {https://doi.org/10.1103/PhysRevLett.123.080502} {\bibfield  {journal}
  {\bibinfo  {journal} {Physical Review Letters}\ }\textbf {\bibinfo {volume}
  {123}},\ \bibinfo {pages} {080502} (\bibinfo {year} {2019})}\BibitemShut
  {NoStop}%
\bibitem [{\citenamefont {Rakonjac}\ \emph {et~al.}(2021)\citenamefont
  {Rakonjac}, \citenamefont {Lago-Rivera}, \citenamefont {Seri}, \citenamefont
  {Mazzera}, \citenamefont {Grandi},\ and\ \citenamefont
  {de~Riedmatten}}]{Rakonjac2021a}%
  \BibitemOpen
  \bibfield  {author} {\bibinfo {author} {\bibfnamefont {J.~V.}\ \bibnamefont
  {Rakonjac}}, \bibinfo {author} {\bibfnamefont {D.}~\bibnamefont
  {Lago-Rivera}}, \bibinfo {author} {\bibfnamefont {A.}~\bibnamefont {Seri}},
  \bibinfo {author} {\bibfnamefont {M.}~\bibnamefont {Mazzera}}, \bibinfo
  {author} {\bibfnamefont {S.}~\bibnamefont {Grandi}},\ and\ \bibinfo {author}
  {\bibfnamefont {H.}~\bibnamefont {de~Riedmatten}},\ }\href
  {https://doi.org/10.1103/PhysRevLett.127.210502} {\bibfield  {journal}
  {\bibinfo  {journal} {Physical Review Letters}\ }\textbf {\bibinfo {volume}
  {127}},\ \bibinfo {pages} {210502} (\bibinfo {year} {2021})}\BibitemShut
  {NoStop}%
\bibitem [{\citenamefont {Saglamyurek}\ \emph {et~al.}(2015)\citenamefont
  {Saglamyurek}, \citenamefont {Jin}, \citenamefont {Verma}, \citenamefont
  {Shaw}, \citenamefont {Marsili}, \citenamefont {Nam}, \citenamefont {Oblak},\
  and\ \citenamefont {Tittel}}]{Saglamyurek2015}%
  \BibitemOpen
  \bibfield  {author} {\bibinfo {author} {\bibfnamefont {E.}~\bibnamefont
  {Saglamyurek}}, \bibinfo {author} {\bibfnamefont {J.}~\bibnamefont {Jin}},
  \bibinfo {author} {\bibfnamefont {V.~B.}\ \bibnamefont {Verma}}, \bibinfo
  {author} {\bibfnamefont {M.~D.}\ \bibnamefont {Shaw}}, \bibinfo {author}
  {\bibfnamefont {F.}~\bibnamefont {Marsili}}, \bibinfo {author} {\bibfnamefont
  {S.~W.}\ \bibnamefont {Nam}}, \bibinfo {author} {\bibfnamefont
  {D.}~\bibnamefont {Oblak}},\ and\ \bibinfo {author} {\bibfnamefont
  {W.}~\bibnamefont {Tittel}},\ }\href
  {https://doi.org/10.1038/nphoton.2014.311} {\bibfield  {journal} {\bibinfo
  {journal} {Nature Photonics}\ }\textbf {\bibinfo {volume} {9}},\ \bibinfo
  {pages} {83} (\bibinfo {year} {2015})}\BibitemShut {NoStop}%
\bibitem [{\citenamefont {Craiciu}\ \emph {et~al.}(2021)\citenamefont
  {Craiciu}, \citenamefont {Lei}, \citenamefont {Rochman}, \citenamefont
  {Bartholomew},\ and\ \citenamefont {Faraon}}]{Craiciu2021}%
  \BibitemOpen
  \bibfield  {author} {\bibinfo {author} {\bibfnamefont {I.}~\bibnamefont
  {Craiciu}}, \bibinfo {author} {\bibfnamefont {M.}~\bibnamefont {Lei}},
  \bibinfo {author} {\bibfnamefont {J.}~\bibnamefont {Rochman}}, \bibinfo
  {author} {\bibfnamefont {J.~G.}\ \bibnamefont {Bartholomew}},\ and\ \bibinfo
  {author} {\bibfnamefont {A.}~\bibnamefont {Faraon}},\ }\href
  {https://doi.org/10.1364/OPTICA.412211} {\bibfield  {journal} {\bibinfo
  {journal} {Optica}\ }\textbf {\bibinfo {volume} {8}},\ \bibinfo {pages} {114}
  (\bibinfo {year} {2021})}\BibitemShut {NoStop}%
\bibitem [{\citenamefont {Corrielli}\ \emph {et~al.}(2016)\citenamefont
  {Corrielli}, \citenamefont {Seri}, \citenamefont {Mazzera}, \citenamefont
  {Osellame},\ and\ \citenamefont {de~Riedmatten}}]{Corrielli2016}%
  \BibitemOpen
  \bibfield  {author} {\bibinfo {author} {\bibfnamefont {G.}~\bibnamefont
  {Corrielli}}, \bibinfo {author} {\bibfnamefont {A.}~\bibnamefont {Seri}},
  \bibinfo {author} {\bibfnamefont {M.}~\bibnamefont {Mazzera}}, \bibinfo
  {author} {\bibfnamefont {R.}~\bibnamefont {Osellame}},\ and\ \bibinfo
  {author} {\bibfnamefont {H.}~\bibnamefont {de~Riedmatten}},\ }\href
  {https://doi.org/10.1103/PhysRevApplied.5.054013} {\bibfield  {journal}
  {\bibinfo  {journal} {Phys. Rev. Applied}\ }\textbf {\bibinfo {volume} {5}},\
  \bibinfo {pages} {054013} (\bibinfo {year} {2016})}\BibitemShut {NoStop}%
\bibitem [{\citenamefont {Zhong}\ \emph {et~al.}(2017)\citenamefont {Zhong},
  \citenamefont {Kindem}, \citenamefont {Bartholomew}, \citenamefont {Rochman},
  \citenamefont {Craiciu}, \citenamefont {Miyazono}, \citenamefont
  {Bettinelli}, \citenamefont {Cavalli}, \citenamefont {Verma}, \citenamefont
  {Nam}, \citenamefont {Marsili}, \citenamefont {Shaw}, \citenamefont {Beyer},\
  and\ \citenamefont {Faraon}}]{Zhong2017}%
  \BibitemOpen
  \bibfield  {author} {\bibinfo {author} {\bibfnamefont {T.}~\bibnamefont
  {Zhong}}, \bibinfo {author} {\bibfnamefont {J.~M.}\ \bibnamefont {Kindem}},
  \bibinfo {author} {\bibfnamefont {J.~G.}\ \bibnamefont {Bartholomew}},
  \bibinfo {author} {\bibfnamefont {J.}~\bibnamefont {Rochman}}, \bibinfo
  {author} {\bibfnamefont {I.}~\bibnamefont {Craiciu}}, \bibinfo {author}
  {\bibfnamefont {E.}~\bibnamefont {Miyazono}}, \bibinfo {author}
  {\bibfnamefont {M.}~\bibnamefont {Bettinelli}}, \bibinfo {author}
  {\bibfnamefont {E.}~\bibnamefont {Cavalli}}, \bibinfo {author} {\bibfnamefont
  {V.}~\bibnamefont {Verma}}, \bibinfo {author} {\bibfnamefont {S.~W.}\
  \bibnamefont {Nam}}, \bibinfo {author} {\bibfnamefont {F.}~\bibnamefont
  {Marsili}}, \bibinfo {author} {\bibfnamefont {M.~D.}\ \bibnamefont {Shaw}},
  \bibinfo {author} {\bibfnamefont {A.~D.}\ \bibnamefont {Beyer}},\ and\
  \bibinfo {author} {\bibfnamefont {A.}~\bibnamefont {Faraon}},\ }\href
  {https://doi.org/10.1126/science.aan5959} {\bibfield  {journal} {\bibinfo
  {journal} {Science}\ }\textbf {\bibinfo {volume} {357}},\ \bibinfo {pages}
  {1392} (\bibinfo {year} {2017})}\BibitemShut {NoStop}%
\bibitem [{\citenamefont {Seri}\ \emph {et~al.}(2018)\citenamefont {Seri},
  \citenamefont {Corrielli}, \citenamefont {Lago-Rivera}, \citenamefont
  {Lenhard}, \citenamefont {de~Riedmatten}, \citenamefont {Osellame},\ and\
  \citenamefont {Mazzera}}]{Seri2018}%
  \BibitemOpen
  \bibfield  {author} {\bibinfo {author} {\bibfnamefont {A.}~\bibnamefont
  {Seri}}, \bibinfo {author} {\bibfnamefont {G.}~\bibnamefont {Corrielli}},
  \bibinfo {author} {\bibfnamefont {D.}~\bibnamefont {Lago-Rivera}}, \bibinfo
  {author} {\bibfnamefont {A.}~\bibnamefont {Lenhard}}, \bibinfo {author}
  {\bibfnamefont {H.}~\bibnamefont {de~Riedmatten}}, \bibinfo {author}
  {\bibfnamefont {R.}~\bibnamefont {Osellame}},\ and\ \bibinfo {author}
  {\bibfnamefont {M.}~\bibnamefont {Mazzera}},\ }\href
  {https://doi.org/10.1364/OPTICA.5.000934} {\bibfield  {journal} {\bibinfo
  {journal} {Optica}\ }\textbf {\bibinfo {volume} {5}},\ \bibinfo {pages} {934}
  (\bibinfo {year} {2018})}\BibitemShut {NoStop}%
\bibitem [{\citenamefont {Liu}\ \emph {et~al.}(2020{\natexlab{a}})\citenamefont
  {Liu}, \citenamefont {Zhu}, \citenamefont {Su}, \citenamefont {Ma},
  \citenamefont {Zhou}, \citenamefont {Li},\ and\ \citenamefont
  {Guo}}]{Liu2020a}%
  \BibitemOpen
  \bibfield  {author} {\bibinfo {author} {\bibfnamefont {C.}~\bibnamefont
  {Liu}}, \bibinfo {author} {\bibfnamefont {T.-X.}\ \bibnamefont {Zhu}},
  \bibinfo {author} {\bibfnamefont {M.-X.}\ \bibnamefont {Su}}, \bibinfo
  {author} {\bibfnamefont {Y.-Z.}\ \bibnamefont {Ma}}, \bibinfo {author}
  {\bibfnamefont {Z.-Q.}\ \bibnamefont {Zhou}}, \bibinfo {author}
  {\bibfnamefont {C.-F.}\ \bibnamefont {Li}},\ and\ \bibinfo {author}
  {\bibfnamefont {G.-C.}\ \bibnamefont {Guo}},\ }\href
  {https://doi.org/10.1103/PhysRevLett.125.260504} {\bibfield  {journal}
  {\bibinfo  {journal} {Physical Review Letters}\ }\textbf {\bibinfo {volume}
  {125}},\ \bibinfo {pages} {260504} (\bibinfo {year}
  {2020}{\natexlab{a}})}\BibitemShut {NoStop}%
\bibitem [{\citenamefont {Dibos}\ \emph {et~al.}(2018)\citenamefont {Dibos},
  \citenamefont {Raha}, \citenamefont {Phenicie},\ and\ \citenamefont
  {Thompson}}]{Dibos2018}%
  \BibitemOpen
  \bibfield  {author} {\bibinfo {author} {\bibfnamefont {A.~M.}\ \bibnamefont
  {Dibos}}, \bibinfo {author} {\bibfnamefont {M.}~\bibnamefont {Raha}},
  \bibinfo {author} {\bibfnamefont {C.~M.}\ \bibnamefont {Phenicie}},\ and\
  \bibinfo {author} {\bibfnamefont {J.~D.}\ \bibnamefont {Thompson}},\ }\href
  {https://doi.org/10.1103/PhysRevLett.120.243601} {\bibfield  {journal}
  {\bibinfo  {journal} {Phys. Rev. Lett.}\ }\textbf {\bibinfo {volume} {120}},\
  \bibinfo {pages} {243601} (\bibinfo {year} {2018})}\BibitemShut {NoStop}%
\bibitem [{\citenamefont {Kindem}\ \emph {et~al.}(2020)\citenamefont {Kindem},
  \citenamefont {Ruskuc}, \citenamefont {Bartholomew}, \citenamefont {Rochman},
  \citenamefont {Huan},\ and\ \citenamefont {Faraon}}]{Kindem2020}%
  \BibitemOpen
  \bibfield  {author} {\bibinfo {author} {\bibfnamefont {J.~M.}\ \bibnamefont
  {Kindem}}, \bibinfo {author} {\bibfnamefont {A.}~\bibnamefont {Ruskuc}},
  \bibinfo {author} {\bibfnamefont {J.~G.}\ \bibnamefont {Bartholomew}},
  \bibinfo {author} {\bibfnamefont {J.}~\bibnamefont {Rochman}}, \bibinfo
  {author} {\bibfnamefont {Y.~Q.}\ \bibnamefont {Huan}},\ and\ \bibinfo
  {author} {\bibfnamefont {A.}~\bibnamefont {Faraon}},\ }\href
  {https://doi.org/10.1038/s41586-020-2160-9} {\bibfield  {journal} {\bibinfo
  {journal} {Nature}\ }\textbf {\bibinfo {volume} {580}},\ \bibinfo {pages}
  {201} (\bibinfo {year} {2020})}\BibitemShut {NoStop}%
\bibitem [{\citenamefont {Saglamyurek}\ \emph {et~al.}(2011)\citenamefont
  {Saglamyurek}, \citenamefont {Sinclair}, \citenamefont {Jin}, \citenamefont
  {Slater}, \citenamefont {Oblak}, \citenamefont {Bussi\`{e}res}, \citenamefont
  {George}, \citenamefont {Ricken}, \citenamefont {Sohler},\ and\ \citenamefont
  {Tittel}}]{Saglamyurek2011}%
  \BibitemOpen
  \bibfield  {author} {\bibinfo {author} {\bibfnamefont {E.}~\bibnamefont
  {Saglamyurek}}, \bibinfo {author} {\bibfnamefont {N.}~\bibnamefont
  {Sinclair}}, \bibinfo {author} {\bibfnamefont {J.}~\bibnamefont {Jin}},
  \bibinfo {author} {\bibfnamefont {J.~A.}\ \bibnamefont {Slater}}, \bibinfo
  {author} {\bibfnamefont {D.}~\bibnamefont {Oblak}}, \bibinfo {author}
  {\bibfnamefont {F.}~\bibnamefont {Bussi\`{e}res}}, \bibinfo {author}
  {\bibfnamefont {M.}~\bibnamefont {George}}, \bibinfo {author} {\bibfnamefont
  {R.}~\bibnamefont {Ricken}}, \bibinfo {author} {\bibfnamefont
  {W.}~\bibnamefont {Sohler}},\ and\ \bibinfo {author} {\bibfnamefont
  {W.}~\bibnamefont {Tittel}},\ }\href {http://dx.doi.org/10.1038/nature09719}
  {\bibfield  {journal} {\bibinfo  {journal} {Nature}\ }\textbf {\bibinfo
  {volume} {469}},\ \bibinfo {pages} {512} (\bibinfo {year}
  {2011})}\BibitemShut {NoStop}%
\bibitem [{Sup()}]{SuppMat}%
  \BibitemOpen
  \href@noop {} {}\bibinfo {note} {See Supplemental Materials.}\BibitemShut
  {Stop}%
\bibitem [{\citenamefont {Afzelius}\ \emph {et~al.}(2009)\citenamefont
  {Afzelius}, \citenamefont {Simon}, \citenamefont {de~Riedmatten},\ and\
  \citenamefont {Gisin}}]{Afzelius2009}%
  \BibitemOpen
  \bibfield  {author} {\bibinfo {author} {\bibfnamefont {M.}~\bibnamefont
  {Afzelius}}, \bibinfo {author} {\bibfnamefont {C.}~\bibnamefont {Simon}},
  \bibinfo {author} {\bibfnamefont {H.}~\bibnamefont {de~Riedmatten}},\ and\
  \bibinfo {author} {\bibfnamefont {N.}~\bibnamefont {Gisin}},\ }\href
  {https://doi.org/10.1103/PhysRevA.79.052329} {\bibfield  {journal} {\bibinfo
  {journal} {Physical Review A}\ }\textbf {\bibinfo {volume} {79}},\ \bibinfo
  {pages} {052329} (\bibinfo {year} {2009})}\BibitemShut {NoStop}%
\bibitem [{\citenamefont {Jobez}\ \emph {et~al.}(2016)\citenamefont {Jobez},
  \citenamefont {Timoney}, \citenamefont {Laplane}, \citenamefont {Etesse},
  \citenamefont {Ferrier}, \citenamefont {Goldner}, \citenamefont {Gisin},\
  and\ \citenamefont {Afzelius}}]{Jobez2016}%
  \BibitemOpen
  \bibfield  {author} {\bibinfo {author} {\bibfnamefont {P.}~\bibnamefont
  {Jobez}}, \bibinfo {author} {\bibfnamefont {N.}~\bibnamefont {Timoney}},
  \bibinfo {author} {\bibfnamefont {C.}~\bibnamefont {Laplane}}, \bibinfo
  {author} {\bibfnamefont {J.}~\bibnamefont {Etesse}}, \bibinfo {author}
  {\bibfnamefont {A.}~\bibnamefont {Ferrier}}, \bibinfo {author} {\bibfnamefont
  {P.}~\bibnamefont {Goldner}}, \bibinfo {author} {\bibfnamefont
  {N.}~\bibnamefont {Gisin}},\ and\ \bibinfo {author} {\bibfnamefont
  {M.}~\bibnamefont {Afzelius}},\ }\href
  {https://doi.org/10.1103/PhysRevA.93.032327} {\bibfield  {journal} {\bibinfo
  {journal} {Physical Review A}\ }\textbf {\bibinfo {volume} {93}},\ \bibinfo
  {pages} {032327} (\bibinfo {year} {2016})}\BibitemShut {NoStop}%
\bibitem [{\citenamefont {Fasel}\ \emph {et~al.}(2004)\citenamefont {Fasel},
  \citenamefont {Alibart}, \citenamefont {Tanzilli}, \citenamefont {Baldi},
  \citenamefont {Beveratos}, \citenamefont {Gisin},\ and\ \citenamefont
  {Zbinden}}]{Fasel2004}%
  \BibitemOpen
  \bibfield  {author} {\bibinfo {author} {\bibfnamefont {S.}~\bibnamefont
  {Fasel}}, \bibinfo {author} {\bibfnamefont {O.}~\bibnamefont {Alibart}},
  \bibinfo {author} {\bibfnamefont {S.}~\bibnamefont {Tanzilli}}, \bibinfo
  {author} {\bibfnamefont {P.}~\bibnamefont {Baldi}}, \bibinfo {author}
  {\bibfnamefont {A.}~\bibnamefont {Beveratos}}, \bibinfo {author}
  {\bibfnamefont {N.}~\bibnamefont {Gisin}},\ and\ \bibinfo {author}
  {\bibfnamefont {H.}~\bibnamefont {Zbinden}},\ }\href
  {https://doi.org/10.1088/1367-2630/6/1/163} {\bibfield  {journal} {\bibinfo
  {journal} {New Journal of Physics}\ }\textbf {\bibinfo {volume} {6}},\
  \bibinfo {pages} {163} (\bibinfo {year} {2004})}\BibitemShut {NoStop}%
\bibitem [{\citenamefont {Franson}(1989)}]{Franson1989}%
  \BibitemOpen
  \bibfield  {author} {\bibinfo {author} {\bibfnamefont {J.~D.}\ \bibnamefont
  {Franson}},\ }\href {https://doi.org/10.1103/PhysRevLett.62.2205} {\bibfield
  {journal} {\bibinfo  {journal} {Physical Review Letters}\ }\textbf {\bibinfo
  {volume} {62}},\ \bibinfo {pages} {2205} (\bibinfo {year}
  {1989})}\BibitemShut {NoStop}%
\bibitem [{\citenamefont {G\"{u}ndo\u{g}an}\ \emph {et~al.}(2015)\citenamefont
  {G\"{u}ndo\u{g}an}, \citenamefont {Ledingham}, \citenamefont {Kutluer},
  \citenamefont {Mazzera},\ and\ \citenamefont {de~Riedmatten}}]{Gundogan2015}%
  \BibitemOpen
  \bibfield  {author} {\bibinfo {author} {\bibfnamefont {M.}~\bibnamefont
  {G\"{u}ndo\u{g}an}}, \bibinfo {author} {\bibfnamefont {P.~M.}\ \bibnamefont
  {Ledingham}}, \bibinfo {author} {\bibfnamefont {K.}~\bibnamefont {Kutluer}},
  \bibinfo {author} {\bibfnamefont {M.}~\bibnamefont {Mazzera}},\ and\ \bibinfo
  {author} {\bibfnamefont {H.}~\bibnamefont {de~Riedmatten}},\ }\href
  {http://link.aps.org/doi/10.1103/PhysRevLett.114.230501} {\bibfield
  {journal} {\bibinfo  {journal} {Physical Review Letters}\ }\textbf {\bibinfo
  {volume} {114}},\ \bibinfo {pages} {230501} (\bibinfo {year}
  {2015})}\BibitemShut {NoStop}%
\bibitem [{\citenamefont {Seri}(2019)}]{Seri2019thesis}%
  \BibitemOpen
  \bibfield  {author} {\bibinfo {author} {\bibfnamefont {A.}~\bibnamefont
  {Seri}},\ }\emph {\bibinfo {title} {{A multimode solid-state quantum memory
  for single photons}}},\ \href {http://hdl.handle.net/2117/166471} {\bibinfo
  {type} {Phd thesis}},\ \bibinfo  {school} {Universitat Polit{\`{e}}cnica de
  Catalunya.} (\bibinfo {year} {2019})\BibitemShut {NoStop}%
\bibitem [{\citenamefont {Liu}\ \emph {et~al.}(2020{\natexlab{b}})\citenamefont
  {Liu}, \citenamefont {Zhou}, \citenamefont {Zhu}, \citenamefont {Zheng},
  \citenamefont {Jin}, \citenamefont {Liu}, \citenamefont {Li}, \citenamefont
  {Huang}, \citenamefont {Ma}, \citenamefont {Tu}, \citenamefont {Yang},
  \citenamefont {Li},\ and\ \citenamefont {Guo}}]{Liu2020b}%
  \BibitemOpen
  \bibfield  {author} {\bibinfo {author} {\bibfnamefont {C.}~\bibnamefont
  {Liu}}, \bibinfo {author} {\bibfnamefont {Z.-Q.}\ \bibnamefont {Zhou}},
  \bibinfo {author} {\bibfnamefont {T.-X.}\ \bibnamefont {Zhu}}, \bibinfo
  {author} {\bibfnamefont {L.}~\bibnamefont {Zheng}}, \bibinfo {author}
  {\bibfnamefont {M.}~\bibnamefont {Jin}}, \bibinfo {author} {\bibfnamefont
  {X.}~\bibnamefont {Liu}}, \bibinfo {author} {\bibfnamefont {P.-Y.}\
  \bibnamefont {Li}}, \bibinfo {author} {\bibfnamefont {J.-Y.}\ \bibnamefont
  {Huang}}, \bibinfo {author} {\bibfnamefont {Y.}~\bibnamefont {Ma}}, \bibinfo
  {author} {\bibfnamefont {T.}~\bibnamefont {Tu}}, \bibinfo {author}
  {\bibfnamefont {T.-S.}\ \bibnamefont {Yang}}, \bibinfo {author}
  {\bibfnamefont {C.-F.}\ \bibnamefont {Li}},\ and\ \bibinfo {author}
  {\bibfnamefont {G.-C.}\ \bibnamefont {Guo}},\ }\href
  {https://doi.org/10.1364/optica.379166} {\bibfield  {journal} {\bibinfo
  {journal} {Optica}\ }\textbf {\bibinfo {volume} {7}},\ \bibinfo {pages} {192}
  (\bibinfo {year} {2020}{\natexlab{b}})}\BibitemShut {NoStop}%
\bibitem [{\citenamefont {Sinclair}\ \emph {et~al.}(2014)\citenamefont
  {Sinclair}, \citenamefont {Saglamyurek}, \citenamefont {Mallahzadeh},
  \citenamefont {Slater}, \citenamefont {George}, \citenamefont {Ricken},
  \citenamefont {Hedges}, \citenamefont {Oblak}, \citenamefont {Simon},
  \citenamefont {Sohler},\ and\ \citenamefont {Tittel}}]{Sinclair2014}%
  \BibitemOpen
  \bibfield  {author} {\bibinfo {author} {\bibfnamefont {N.}~\bibnamefont
  {Sinclair}}, \bibinfo {author} {\bibfnamefont {E.}~\bibnamefont
  {Saglamyurek}}, \bibinfo {author} {\bibfnamefont {H.}~\bibnamefont
  {Mallahzadeh}}, \bibinfo {author} {\bibfnamefont {J.~A.}\ \bibnamefont
  {Slater}}, \bibinfo {author} {\bibfnamefont {M.}~\bibnamefont {George}},
  \bibinfo {author} {\bibfnamefont {R.}~\bibnamefont {Ricken}}, \bibinfo
  {author} {\bibfnamefont {M.~P.}\ \bibnamefont {Hedges}}, \bibinfo {author}
  {\bibfnamefont {D.}~\bibnamefont {Oblak}}, \bibinfo {author} {\bibfnamefont
  {C.}~\bibnamefont {Simon}}, \bibinfo {author} {\bibfnamefont
  {W.}~\bibnamefont {Sohler}},\ and\ \bibinfo {author} {\bibfnamefont
  {W.}~\bibnamefont {Tittel}},\ }\href
  {http://link.aps.org/doi/10.1103/PhysRevLett.113.053603} {\bibfield
  {journal} {\bibinfo  {journal} {Physical Review Letters}\ }\textbf {\bibinfo
  {volume} {113}},\ \bibinfo {pages} {053603} (\bibinfo {year}
  {2014})}\BibitemShut {NoStop}%
\bibitem [{\citenamefont {Laplane}\ \emph {et~al.}(2017)\citenamefont
  {Laplane}, \citenamefont {Jobez}, \citenamefont {Etesse}, \citenamefont
  {Gisin},\ and\ \citenamefont {Afzelius}}]{Laplane2017}%
  \BibitemOpen
  \bibfield  {author} {\bibinfo {author} {\bibfnamefont {C.}~\bibnamefont
  {Laplane}}, \bibinfo {author} {\bibfnamefont {P.}~\bibnamefont {Jobez}},
  \bibinfo {author} {\bibfnamefont {J.}~\bibnamefont {Etesse}}, \bibinfo
  {author} {\bibfnamefont {N.}~\bibnamefont {Gisin}},\ and\ \bibinfo {author}
  {\bibfnamefont {M.}~\bibnamefont {Afzelius}},\ }\href
  {https://doi.org/10.1103/PhysRevLett.118.210501} {\bibfield  {journal}
  {\bibinfo  {journal} {Physical Review Letters}\ }\textbf {\bibinfo {volume}
  {118}},\ \bibinfo {pages} {210501} (\bibinfo {year} {2017})}\BibitemShut
  {NoStop}%
\bibitem [{\citenamefont {Ma}\ \emph {et~al.}(2021)\citenamefont {Ma},
  \citenamefont {Ma}, \citenamefont {Zhou}, \citenamefont {Li},\ and\
  \citenamefont {Guo}}]{Ma2021}%
  \BibitemOpen
  \bibfield  {author} {\bibinfo {author} {\bibfnamefont {Y.}~\bibnamefont
  {Ma}}, \bibinfo {author} {\bibfnamefont {Y.-Z.}\ \bibnamefont {Ma}}, \bibinfo
  {author} {\bibfnamefont {Z.-Q.}\ \bibnamefont {Zhou}}, \bibinfo {author}
  {\bibfnamefont {C.-F.}\ \bibnamefont {Li}},\ and\ \bibinfo {author}
  {\bibfnamefont {G.-C.}\ \bibnamefont {Guo}},\ }\href
  {https://doi.org/10.1038/s41467-021-22706-y} {\bibfield  {journal} {\bibinfo
  {journal} {Nature Communications}\ }\textbf {\bibinfo {volume} {12}},\
  \bibinfo {pages} {2381} (\bibinfo {year} {2021})}\BibitemShut {NoStop}%
\bibitem [{\citenamefont {Ortu}\ \emph {et~al.}(2021)\citenamefont {Ortu},
  \citenamefont {Holz{\"{a}}pfel}, \citenamefont {Etesse},\ and\ \citenamefont
  {Afzelius}}]{Ortu2021}%
  \BibitemOpen
  \bibfield  {author} {\bibinfo {author} {\bibfnamefont {A.}~\bibnamefont
  {Ortu}}, \bibinfo {author} {\bibfnamefont {A.}~\bibnamefont
  {Holz{\"{a}}pfel}}, \bibinfo {author} {\bibfnamefont {J.}~\bibnamefont
  {Etesse}},\ and\ \bibinfo {author} {\bibfnamefont {M.}~\bibnamefont
  {Afzelius}},\ }\href {http://arxiv.org/abs/2109.06669} {\bibfield  {journal}
  {\bibinfo  {journal} {arxiv:2109.06669}\ } (\bibinfo {year}
  {2021})}\BibitemShut {NoStop}%
\end{thebibliography}
\end{document}